\documentclass{fundam}

\usepackage[english]{babel}
\usepackage[nottoc]{tocbibind}
\usepackage{graphicx}
\usepackage{caption}
\usepackage{hyperref}

\begin{document}


\setcounter{page}{171}
\publyear{22}
\papernumber{2157}
\volume{189}
\issue{2}

\finalVersionForARXIV


\runninghead{D. Coluzzi and G. Baselli}{Diffuse and Localized Functional Dysconnectivity in Schizophrenia}

\title{Diffuse and Localized Functional Dysconnectivity in Schizophrenia: a Bootstrapped Top-Down Approach}

\author{Davide Coluzzi\thanks{Address for  correspondence: Dipartimento di Elettronica,
                       Informazione e Bioingegneria, Politecnico di Milano, Italy}, \, Giuseppe Baselli
\\
Dipartimento di Elettronica, Informazione e Bioingegneria\\
Politecnico di Milano, Italy \\
davide.coluzzi@polimi.it,  giuseppe.baselli@polimi.it
}

\maketitle

\vspace*{-5mm}
\begin{abstract}
Schizophrenia is a brain disorder leading to detached mind's normally integrated processes. Hence, the exploration of the symptoms in relation to functional connectivity (FC) had great relevance in the field. Connectivity can be investigated on different levels, going from global features to single edges between pairs of regions, revealing diffuse and localized dysconnection patterns. In this context, schizophrenia is characterized by a different global integration with reduced connectivity in specific areas of the brain, part of the Default Mode Network (DMN). However, the assessment of FC presents various sources of uncertainty. This study proposes a multi-level approach for more robust group-comparison. \\
FC data between 74 AAL brain areas of 15 healthy controls (HC) and 12 subjects with chronic schizophrenia (SZ) were used. Multi-level analyses were carried out by the previously published SPIDER-NET tool. Graph topological indexes were evaluated to assess global abnormalities. Robustness was augmented by bootstrapped (BOOT) data and the stability was evaluated by removing one (RST1) or two subjects (RST2). The DMN subgraph was extracted and specifically evaluated. Changes relevant to the overall local indexes were also analyzed. Finally, the connection weights were explored to enhance common strongest activations/deactivations.\\
At a global level, expected trends of the indexes were found and the significance of modularity ($p=0.043$) was not confirmed by BOOT ($p=0.133$). The robustness assessment tests (both RST1 and RST2) highlighted more stable results for BOOT compared to the direct data testing. Conversely, significant results were found in the analysis at lower levels. The DMN highlighted reduced connectivity and strength as well as increased deactivation in the SZ group. At local level, 13 areas were found to be significantly different ($p<0.05$) in the groups, highlighting a greater divergence in the frontal lobe. These results were confirmed analyzing the single negative edges, suggesting inverted connectivity between prefronto-temporal areas. \\
In conclusion, multi-level analysis supported by BOOT is highly recommended when analyzing FC, especially when diffuse and localized dysconnections must be investigated in limited samples.
\end{abstract}

\section{Introduction}\label{introduction}

The brain is the most tangled network known in nature. Indeed, it comprises billions of neurons which form trillions of synapses between each other. Conversely, at a macroscale extent, brain regions can be thought as connected through fiber bundles, mirroring the anatomical structure, or functional activations of the areas, in resting-state or during a task. All this collective activity makes behaviour, thought, memory and consciousness possible. In this context, clinical disorders are characterized by alterations of the connections’ paths. For this reason, it is paramount to explore these abnormalities to enforce our understanding of the brain in health and disease.

Considering the methods used to investigate brain networks, Magnetic Resonance Imaging (MRI) is the dominant technique for macroscale analysis mainly because of its safety, spatial resolution and availability. On one hand, Diffusion Tensor Imaging (DTI) or other methods allow to visualize and examine the organization of structural connectivity by white matter (WM) tracts. On the other hand, functional MRI (fMRI) inspects the dynamics of activity in each gray matter (GM) area. The functional activations of the GM areas are based on the Blood Oxygen Level Dependent (BOLD) response.

Through these methodologies it is possible to obtain connectivity matrices, which can be investigated in their organization and function. Thus, a matrix (graph) can be defined as a collection of nodes (brain regions) and edges (anatomical or functional connections) between pairs of nodes \cite{Rubinov2011Weight-conserving}. The nodes are obtained by mapping the parcels of the network according to well-known brain atlases defined from the structural or functional point of view \cite{Desikan2006automated,Destrieux2010Automatic}. Accordingly, the edges represent the links detected between the regions describing the weights of the connections. Indeed, the strength of each connection is weighted according to the addressed connectivity measure such as the number of fibers in the structural case, or the Pearson's correlation in the functional case.

The comprehension of these networks is a complex field of research named connectomics, which can potentially address the brain at all its scales, levels and features \cite{Fornito2016Fundamentals}. Focusing on the macro-scale of brain GM areas, it is first possible to analyze and compare different networks through the investigation of their graph-based global (top level) topological features such as node degree, node strength, clustering coefficient, efficiency, path length or modularity \cite{Rubinov2011Weight-conserving,Colon-Perez2016Small}.

Second, the brain is also known to be divided into interacting specialized sub-networks, devoted to specific domains of behavior and cognition \cite{Bassett2017Network}. Extracting sub-networks is an increasingly common practice in explorative studies to easily interpret the brain connectivity data in physiological and pathological conditions \cite{Bassett2017Network, Zalesky2010Network-based, Berron2020Medial, Isernia2021Resting-State, Bordier2018Disrupted,Coluzzi}. In this regard, we presented in a previous study a novel software tool allowing the interactive and flexible analysis of anatomical/functional networks and sub-networks called SPIDER-NET \cite{Coluzzi2022Development}.

Third, the above-mentioned graph indexes can also be computed at local level providing information regarding the configuration of the single nodes. Valuable information is extracted about hubs of the networks and regions associated to specific disruptions, such as lesions \cite{Coluzzi2022Development, Cheng2019Altered}.

These three levels of analysis (network, sub-network and node-level) together with the analysis of the single edges resulted helpful in the study of many pathological conditions. Indeed, global changes of relevant properties, different sub-network configurations, disrupted node connectivity and missing relevant connections were already found in stroke, Alzheimer’s disease and other pathologies \cite{Daianu2013Alzheimer's, Crofts2011Network}. However, the integration and the interpretation of the results extracted at all three levels is still difficult. The parallel multi-level analysis can be particularly helpful in case of pathologies, such as the schizophrenia which the localized/diffuse origin is unknown yet \cite{Fornito2012Schizophrenia}.

Schizophrenia may be characterized as a disorder of brain connectivity since it leads to detached mind's processes. The main symptoms of the pathology such as reduced cognitive and emotional deficits acknowledge this breakdown \cite{Fornito2012Schizophrenia}. More specifically, the clinical manifestations of the disorder can be hallucinations, disorganized thinking and agitation, social isolation, emotional flattening, anhedonia and apathy. Often, these manifestations may be preceded by a series of prodromal symptoms, such as inability to perform one's job or neglect of personal hygiene.

Many studies \cite{Liu2008Disrupted, Alexander-Bloch2010Disrupted} explored the connections between these symptoms and functional associations. Specifically, connectivity deficits were found on all abovementioned levels \cite{Fornito2012Schizophrenia}. First, a global connectivity reduction was discovered in different studies \cite{Alexander-Bloch2010Disrupted, Lynall2010Functional, Fornito2011General}. Second, at interconnected sub-network level, fronto-temporal and occipito-temporal dysconnections were found through the use of the network-based statistics \cite{Zalesky2010Network-based}. Third, amongst all brain regions, the prefrontal cortex resulted to be one of the most affected nodes \cite{Liu2008Disrupted, Lynall2010Functional, Fornito2011General}. More generally, Liu and colleagues found a significant alteration of the pattern of small-world topological properties in many brain regions of the prefrontal, parietal and temporal lobes. Four, Skudlarski and colleagues analyzed single connectivity edges, finding differences in the connectivity pattern originating from posterior cingulate cortex within the Default Mode Network (DMN) suggesting to have the most affected connections by the functional reorganization of the schizophrenia \cite{Skudlarski2010Brain}.

DMN is reasonably important in the context of schizophrenia since it is deeply involved in social behavior, control of the emotional state of the individual \cite{Hu2017Review, Godwin2017Functional}, which characterize altered integrated processes by the pathology. From different studies it resulted that connections within this circuit were different than healthy controls, being related to emotion control and memory \cite{Hu2017Review, Zhou2018Altered}. However, the observations for the DMN remained controversial, since the average connectivity of the patients showed higher or lower values than healthy controls according to the dynamic mental state and the individual connections were subject to reorganization \cite{Skudlarski2010Brain}.

Indeed, although the analysis of changes in functional connectivity is a powerful tool to analyze brain organization, due to the limitations of the fMRI techniques, the statistical investigation of brain connectivity datasets is often subject to uncertainty. Indeed, stationarity is often assumed in the interpretation of the results coming from this technique. However, considering the known dynamic and condition-dependent nature of brain activity, it is obvious that the functional connectivity metrics such as the Pearson correlation coefficient will change over time \cite{Hutchison2013Dynamic, Kottaram2019Brain, Liegeois2017Interpreting}. This variability allows to define the paradigm of dynamic functional connectivity analysis, which cannot be ignored, since it also varies within the same subject and even between time windows within the same session. In this context, the evaluation of functional static connectivity is subjected to uncertainty. In addition, functional connectivity acquisitions are characterized by low signal-to-noise ratio (SNR) and non-neural noise related to cardiac and respiratory processes and hardware instability. Finally, other definition issues such as the window length and other confounds remain
  contro- \linebreak versial~\cite{Hindriks2016Can, Leonardi2015On, Lindquist2014Evaluating, Zalesky2015Towards}.

Among all statistical methodologies, bootstrapping was already used in some imaging problems, such as in the work conducted by Lazar and Alexander \cite{Lazar2005Bootstrap}, since it allows to robustly estimate the statistical features of a population from a limited number of measurement samples, without any assumptions about the distribution assumed by the initial data. This approach also represents an alternative to traditional hypothesis testing, since it does not require to have a test statistic satisfying certain assumptions largely dependent on the experimental design and to know the properties of the data. As a result, the main advantage of the method is that the uncertainty variability of the estimator can be quantified, characterizing the dispersion and other errors in the null hypothesis \cite{Kulesa2015Sampling, Gel2017Bootstrap}. Considering the number of bootstrap samples used, the resulting statistics represent a random sample with replacements from the initial distribution characterize by a smaller size. It is worth remarking that obtaining thousands of bootstrap observations from the initial data is not the same as collecting new data. Indeed, the approach is based on an ensemble of simulated data (surrogates) and the usefulness of bootstrapping is related to the quantification of statistical quantities such as the standard error, a possible bias and a confidence interval of particular sample of data.

In the context of investigating graph-structured data, the quantification of the uncertainty intrinsic to the data is essential for their scientific usefulness. In the recent probabilistic study conducted by Green and Shalizi \cite{Green2022Bootstrapping} the bootstrap was applied on simulated random graphs. It was seen that the resampling of the synthetic data was able to approximate the distributions of motif densities, such as, the number of times fixed subgraph appear in the random network. In this work, the bootstrap was also applied to quantify the uncertainty of the network metrics. Another example is given by the study conducted by Gel and colleagues, where bootstrap was applied for the quantification of uncertainty in graph degree distributions of collaboration networks, where for example articles are represented as nodes and cross-references as edges \cite{Gel2017Bootstrap}. Indeed, when applying the bootstrap over a distribution of data, the expected result is a better definition of the variability centered with respect to the initial mean of the same distribution at the increasing of the number of resampled data \cite{Picheny2010Application}. However, a limitation of this approach is related to computational costs since the bootstrapping methods are based on multiple resampling of the original dataset that can be computationally expensive, especially for large datasets.

However, the bootstrapping approach was only partially applied in brain connectivity studies. For example, in the study conducted by Wei and colleagues this approach was used to perform connectivity matrix feature selection in a regression task cognitive traits prediction \cite{Wei2020Bootstrapping}. Spearman correlation analysis was indeed performed between connectivity and cognitive measures in each resample subset to extract a feature vector. In a previous study \cite{Berman2016Levodopa}, the results from mean functional connectivity of Parkinson’s disease patients were qualitatively analyzed through bootstrap.

To the best of our knowledge, there is significant potential for effectively and more quantitatively applying bootstrap techniques on connectivity data, although fundamental issues still need to be addressed. Indeed, connectivity measurements are well-known to be affected by different sources of noise, which can have a strong impact especially with limited population size. Also, the static evaluation of functional connectivity is limited with respect to dynamic functional connectivity \cite{Hutchison2013Dynamic, Leonardi2015On}. In this sense, it is crucial to evaluate the robustness and the stability of potential biomarkers of brain connectivity to improve their usability and understanding. Moreover, there is the need to evaluate abnormalities on multiple levels, especially in conditions characterized by both localized and diffuse degeneration, such as schizophrenia.

Hence, the main aim of this study was the robust assessment of connectivity indexes on different levels through the bootstrap procedure to ensure a reliable detection of abnormalities in schizophrenia. To investigate both diffuse and localized alterations, a top-down analysis of: i) global connectivity deficits, ii) sub-network disruptions, extracting the DMN, iii) dysconnectivity  in individual regions and its inter-subject variability, and iv) abnormalities in positive/negative connections resulting in activation/deactivation circuits or communities is proposed. The methodology was developed in the context of the SPIDER-NET tool \cite{Coluzzi2022Development} to allow automatic multi-level investigations in group studies. Afterwards, the stability of results is assessed by comparing bootstrapping to direct testing via a leave-n-subject-out approach to evaluate the impact of sources of uncertainty in functional connectivity. The whole pipeline of multi-level analysis was validated on a dataset acquired from healthy and schizophrenic subjects highlighting several abnormalities at the examined levels.

\section{Methods}\label{methods}

\subsection{Data acquisition and study population} \label{dataacqstudypop}

Data used in this study were collected in the context of the work conducted by Zalesky and colleagues \cite{Zalesky2010Network-based}. Regarding the written informed consent and ethical approval, reference is made to the original work. The anonymized connectivity dataset is publicly available at https://nitrc.org/projects/nbs/.
The dataset is composed by 15 healthy controls - HC (mean age 33.3 years, $\sigma$ = 9.2 years, 14 males) and 12 subjects with chronic schizophrenia - SZ (mean age 32.8 years, $\sigma$ = 9.2 years, 10 males). The patients were diagnosed according to standard operational criteria in the Diagnostic and Statistical Manual of Mental Disorders IV (American Psychiatric Association, 2000). The matching of the two populations was done according to age, pre-onset IQ and years of education. SZ subjects did not receive medication on the day of acquisition to reduce acute drug effects on the data. T2*-weighted echo-planar images depicting blood oxygenation level dependent contrast were acquired through a 1.5 Tesla scanner (GE Signa, General Electric, Milwaukee, WI).
As regards the matrix construction: the nodes were defined according to a subset of areas of the AAL atlas, whereas the edges were obtained computing the correlation between times series which were previously preprocessed. More specifically, scale 3 of the wavelet transform ($0.03<f<0.06$ Hz) was considered and the filtered time series were corrected for fluctuations of signals through linear regression against reference time courses extracted from seed regions. Some AAL nodes were excluded since it was not possible to correctly estimate the node-averaged time series, due to poor coverage in some subjects. The resulting connectivity matrices for each subject have dimensions $74 \times 74$.  More details on ethical approval, imaging parameters and processing are reported in the cited work \cite{Zalesky2010Network-based}.

Generally, after composing the matrix, a threshold is defined to emphasize its topological features by removing spurious associations. Among all methods, absolute and density thresholding are the most common in connectivity studies \cite{Colon-Perez2016Small, Heuvel2017Proportional}. In this case, we decided to not threshold the matrices since it introduces a confound on properties of the graphs in the context of a group study. On one hand, different levels of sparsity can be obtained by applying the same threshold to all matrices. In this case, it is thus not possible to rule out systematic density variations as the main reason of group abnormalities of graph-based metrics. On the other hand, matrices can be matched according to sparsity using density thresholding, hence selecting a distinct threshold value. In this case, whether the weight distribution in SZ is reduced, the application of density thresholding can strongly affect weighted indexes results \cite{Fornito2011General, Sheffield2016Cognition}. Furthermore, no thresholding allowed to work on all raw connectivity measures, avoiding the loss of information of the original data and assessing the results of the bootstrap methodology in relation to all sources of uncertainty. On one hand, all weakest and thus possible spurious connections were maintained to account them in the robust evaluation of the indexes through bootstrapping. On the other hand, the density differences are only given by the deletion of the negative connections, also allowing the investigation of binary topological indexes. Hence, two versions of same matrix were obtained to analyze separately positive/negative edges and, thus, activations and deactivations.

\begin{figure}[!h]
\vspace*{-3mm}
\centering
\includegraphics[width=0.67\textwidth]{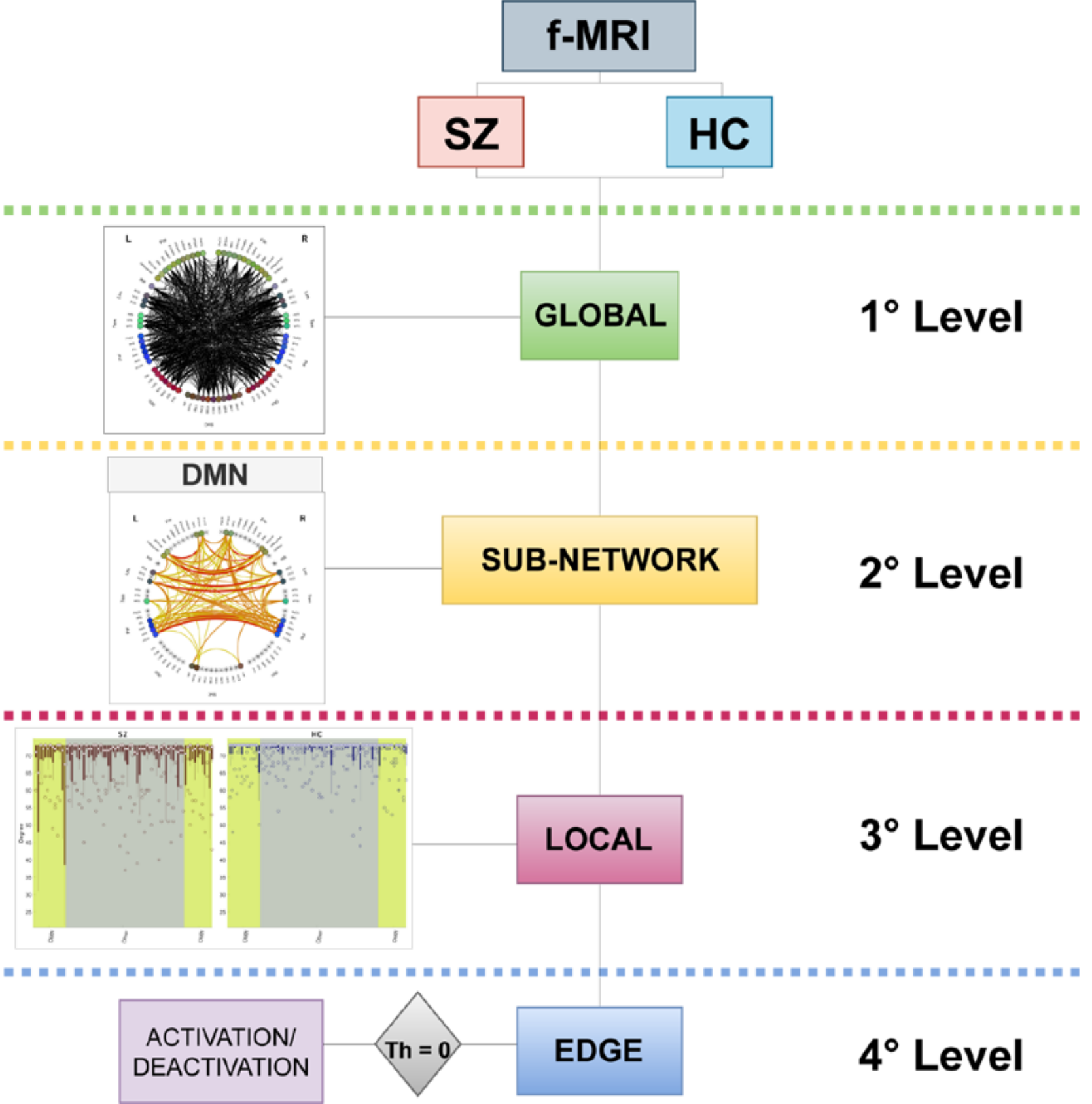}
\captionsetup{justification=justified}
\caption{{Pipeline of the top-down approach proposed. The functional connectivity matrices of the two populations were analyzed from the global level, extracting the DMN as sub-network of interest and investigating the distributions of the local indexes to the edge level, assessing correlations and anticorrelations.{\label{fig1}}%
}}
\end{figure}

\subsection{Top-down bootstrapping approach for group comparison }\label{topdown}

The top-down bootstrapping approach to enable the HC vs SZ group comparison consists of different steps of analysis, done on different levels. The analysis pipeline was built starting from the top level, related to the global characteristics of the matrices coming from the two populations, until the bottom level, related to the single connections. The procedure is summarized in Figure~\ref{fig1}.

First, the global topological indexes were obtained to evaluate the abnormalities in the network topology and global deficits. In order to enable a robust investigation, the values of these properties were bootstrapped, as reported in Section~\ref{statsrob}. Second, bootstrapping was also applied on the extracted sub-network of interest (DMN) and local indexes to visualize significant changes between groups and different subjects of the same groups (see Section~\ref{local}). Finally, the negative connections of the group mean matrixes were analyzed in Section~\ref{connlev}. In this section, the analysis of the communities was performed to enhance common deactivations. Afterwards, a qualitative investigation of strongest and most frequent negative connections reflecting anticorrelations was performed.

The group comparison functions required by the present study were implemented as upgrade of our previously developed SPIDER-NET tool, freely available to researchers \cite{Coluzzi2022Development} The current upgrades will be included in a new release.

\begin{figure}[!b]
\vspace*{-3mm}
\centering
\includegraphics[width=0.85\textwidth]{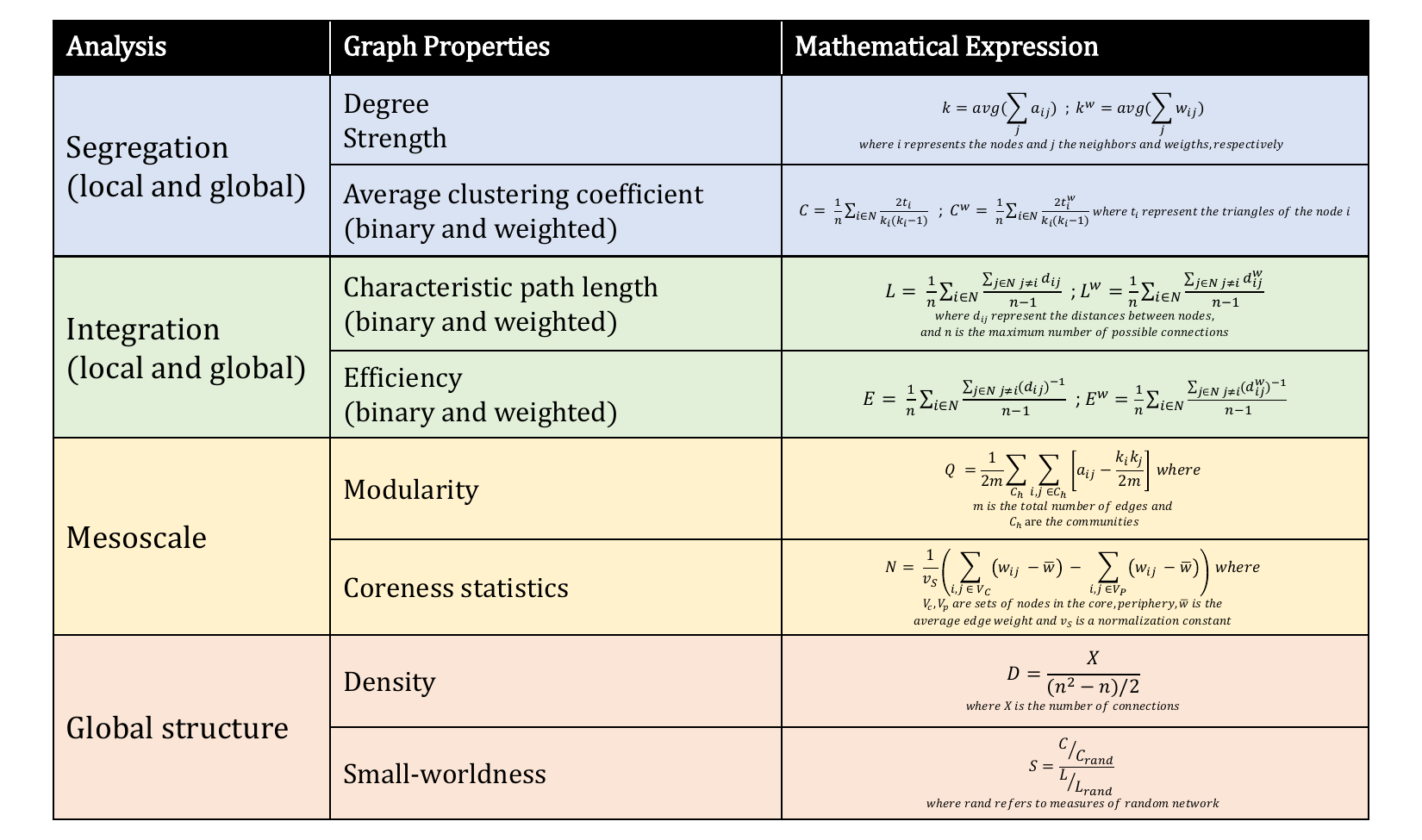}
\caption{{Table of all graph properties considered, divided according to the level of analysis. The segregation and integration properties can be globally assessed through the average of all local values. The mathematical expressions reported indicate global case formulation.{\label{fig2}}%
}}\vspace*{-2mm}
\end{figure}
\subsubsection{Global connectivity}\label{global}

The global topological indexes from the graph theory which were considered are reported in the Figure~\ref{fig2}.

The indexes are divided according to integration/segregation indexes (e.g. path length, clustering coefficient), mesoscale analysis (e.g. modularity, rich-club coefficient) and related to whole network structure organization (e.g. small-worldness).

The degree/strength represent the sum of edges/weights connected to the node. The clustering coefficient quantifies local edge density by counting the triangles average. A triangle occurs if a neighbor of the node is also a neighbor of another neighbor of the node. In the weighted case, the number of triangles is replaced with the geometric mean of its weights. The characteristic path length is a measure of integration which expresses the average shortest path between node pairs. The global efficiency is the average inverse of the characteristic path length, and it represents how efficiently the information travels through the network. All the segregation and integration metrics can be local and global, simply by averaging the values across all nodes.
Mesoscale analysis reveals how much the network present a particular structure. Specifically, the modularity computed through Newman’s approach \cite{Newman2006Modularity} quantifies to what extent the intra-/inter-community link densities are anomalous in comparison to chance. Large values typically reveal significant community structures. Maximized coreness statistic is a measure of how much the network follows the core/periphery paradigm, which is a partition of the network into two groups, where the number and weight of the edges is maximized in the core and minimized within the periphery.

The density is a measure of sparsity of the network. It is the ratio between the number of actual connections and the maximum number of possible connections. A small-world network is a structure in which most nodes are not neighbors of one another. Conversely, the neighbors of any given node are likely to be neighbors of each other, resulting in an easy access by most nodes to every other node with a small number of steps. Small-world networks associate short path length and high clustering coefficient.\vspace*{-1mm}

\paragraph{Statistical Tests and Robustness Assessment}\label{statsrob}

Mann-Whitney tests (MW) were performed to assess statistically significant differences in the global indexes between the two groups. Because of the dimensions of the sample (\#HC = 15; \#SZ = 12), the intrinsic uncertainty given by the non-stationarity of the data \cite{Hutchison2013Dynamic} and the possible presence of spurious connections due to the limitations in the fMRI processing \cite{Leonardi2015On}, which can significantly affect the results, the reliability of significant group differences was also tested.
First, bootstrap hypothesis testing (BOOT) was performed on all global indexes and compared to MW testing. BOOT allows for a better estimation of the null distribution of network measures, providing confidence intervals to evaluate the uncertainty of the statistics and more robust abnormalities detection \cite{Efron1993introduction}. The procedure to test bootstrap hypothesis was the following:

\smallskip
The procedure to test bootstrap hypothesis was the following:
\begin{enumerate}
\item Calculation of the test statistic $\hat{\theta}=|\tilde{x}-\tilde{y}|$, given $x_1,\ldots,x_N$ a random sample from distribution $F$ with median $\tilde{x}$ and $y_1,\ldots,y_M$ another independent random sample from distribution $G$ with median $\tilde{y}$.
\item Bootstrapping: extraction of $B$ sets of random samples $x^*$ (size $N$) and $y^*$ (size $M$) with replacement from $x$ and $y$, respectively.
\eject
\item Calculation of the test statistic $\hat{\theta}^*_b=|\tilde{x}^*_b-\tilde{y}^*_b|$ for each resample.
\item These $B$ resampled test statistics are then made into a null distribution by $\hat{\theta}'_b=\hat{\theta}^*_b-\hat{\theta}$.
\item Estimate of the p-value as
\begin{equation*}
p=\frac{\sum_{b=1}^{B} \left[C\left(\hat{\theta}_{b}^{\prime} \geq \hat{\theta}\right)+1\right]}{B+1}
\end{equation*}
where $C\{condition\}=1$ when the condition is true and $0$ otherwise.
\end{enumerate}

F and G represent the distributions of a global index in HC and SZ, thus N and M representing the number of HC and SZ, 15 and 12 respectively. The resampling from these two sets is carried out 5000 ($B$ - number of resamples) times for all bootstrap hypothesis tests and the test statistic is the difference in medians \cite{Johnston2021bootstrap}.

Second, the robustness of the measures was assessed through a leave-n-subject-out approach, with $n$ equal to 1 and 2. The variations of the statistics computed through both methods were analyzed changing the population samples according to the two procedures.

\begin{figure}[htbp]
\vspace{2mm}
\centering
\hspace*{-2mm}\includegraphics[width=1.01\textwidth]{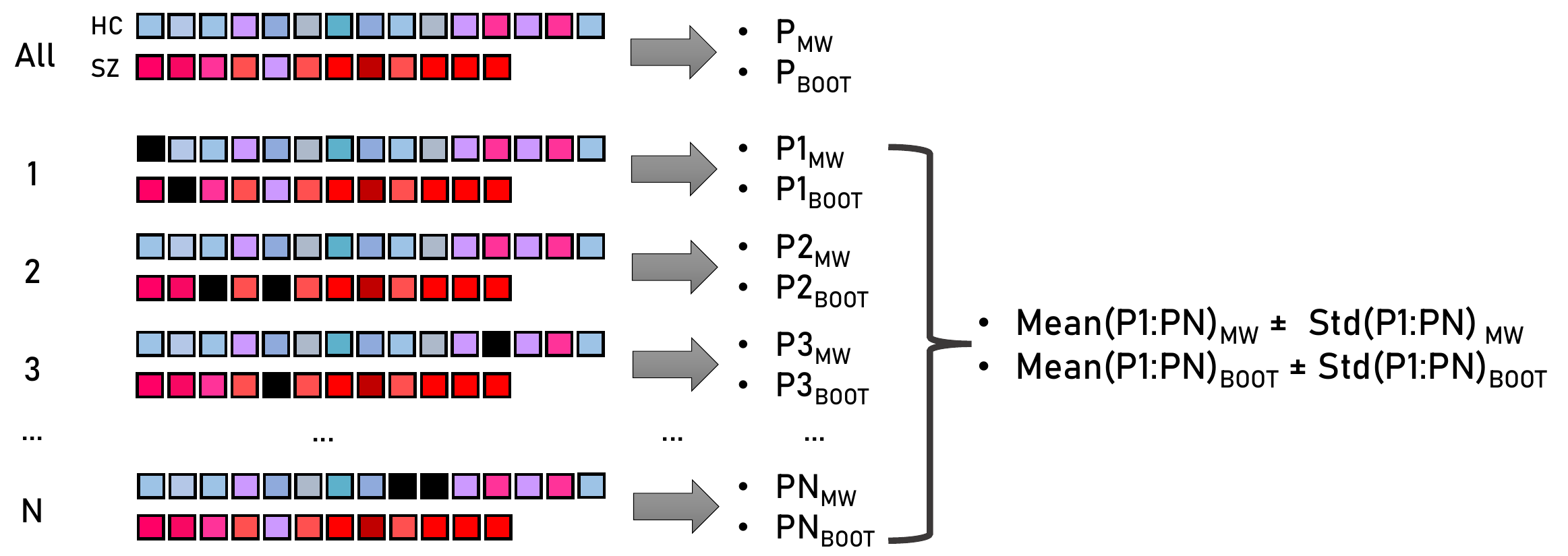}
 \caption{{Schema representing the robustness statistical test randomly removing two subjects (black squares) at each iteration (RST2). In this case, $N$ is equal to 350.{\label{fig3}}%
}}
\end{figure}

The first one, hereafter referred as RST1, evaluates the variation of the p-values obtained removing all subjects one at a time from the populations values to perform MW tests and to create new resamples for BOOT. The second one, hereafter referred as RST2, evaluates the variation of the p-values obtained randomly removing pairs of subjects from the populations values to perform both tests. In this case, 350 random extractions of pairs of subjects were performed. The total possible pairs which can be extracted are $27*26=702$, thus we decided to run the test for about 50\% of the possibilities. RST2 is summarized in Figure~\ref{fig3}.

Then, mean and standard deviation across all tests were evaluated for both procedures.

\subsubsection{Local topological properties and sub-network extraction}\label{local}

Afterwards, local topological properties related to the specific nodes of the network across all subjects in the two groups were analyzed. These indexes are the local counterpart of the global ones, as reported in the Section~\ref{global} and Figure~\ref{fig2}. In this regard, we obtained the distributions of these values assessing differences in the local structure and in the variability across the two groups. Specifically, we focused on degree and strength. The former helps to represent the influence of negative connections, being the original matrixes complete and divided by using a threshold in 0. The latter instead identifies node or set of nodes whose weighted values differ in the two populations. In this case, we performed MW and BOOT, as reported in the previous section~\ref{statsrob}, on the local index values. The resampling was performed with 5000 samples for all BOOT tests.
Furthermore, these differences were investigated across the nodes of the different lobes and the nodes within/outside the DMN to obtain the most deviating local values between the two groups. The distributions per areas and the inter-subject variability, which can be related to the different intra-group characteristics, were \linebreak highlighted.

In order to perform these analyses, we divided the nodes according to the lobes and we extracted the DMN from all matrices of both groups. Specifically, we defined the DMN according to previous studies highlighting the regions characterizing the network \cite{Mevel2011Default, Li2014default, Manning2016Chapter, Alves2019improved, Heuvel2010Exploring,Finotelli}. Generally, the brain regions found to be activated within the DMN comprise the medial frontal cortex, the medial temporal lobe, the posterior cingulate cortex, the ventral precuneus and the inferior parietal cortical regions. Apart from these typical core areas, lateral temporal cortex, hippocampal formation and amygdala are also often reported as parts of the network \cite{Li2014default, Alves2019improved} Thus, from the typical Broadmann areas comprising the DMN (BA: 9, 10m, 10r, 10p, 23/31, 24, 29/30, 32a, 32c, 39, 49) and these additional regions, a list of 13 AAL parcels from the left hemisphere and 12 AAL parcels from the right hemisphere was obtained. The parcel which is present only in the left hemisphere is the amygdala since the right counterpart was not present in the data, due to the previously described issue of poor coverage in some subjects (see Section~\ref{dataacqstudypop}) \cite{Zalesky2010Network-based}. The sub-networks were then composed of these nodes and the edges existing between pairs of DMN areas (“Extract” mode in SPIDER-NET), maintaining the intra-lobe connections.

\subsubsection{Connection-level investigation}\label{connlev}

The examination of the single connections was performed computing the mean matrix of the groups which were compared to visualize differences in weights and deactivation. Thus, the average networks composed by all negative contributions were obtained. Then, community detection analysis through the Newman’s method \cite{Newman2006Modularity} implemented in the Brain Connectivity Toolbox \cite{Rubinov2010Complex} and embedded into SPIDER-NET \cite{Coluzzi2022Development} was performed. The resulting negative sub-networks were obtained and analyzed. In addition, connectograms were drawn considering the entire network with no thresholding apart the separation of positive and negative edges. The rationale was to analyze graphs formed by all brain regions and edges, or all the  DMN areas and edges. Thresholding was applied only for graphical purposes in some connectogram figures where density obscured the main connections, as highlighted in the captions.

\section{Results}\label{results}

First, the global topological properties were assessed. Table~\ref{table} summarizes the results obtained from the comparison between HC and SZ of the overall indexes using nonparametric test and bootstrap hypothesis testing. The results from the robustness tests are also summarized in Table~\ref{table}.

\begin{table}[!h]
\centering
\small
\caption{Results of the statistical tests performed on global indexes.}
\label{tab:graph_indices_comparison}\vspace*{-1.6mm}
\scalebox{0.85}{
\hspace*{-2mm}\begin{tabular}{p{1.5cm}|p{1.65cm} p{2cm} | p{0.35cm} p{0.65cm}|p{1.4cm}p{1.8cm}|p{1.4cm}p{1.6cm}}
 \textbf{Graph Index} & \multicolumn{4}{c|}{\textbf{Whole Comparison}} & \multicolumn{2}{c|}{\textbf{RST1 (\#Sub=1)}} & \multicolumn{2}{c}{\textbf{RST2 (\#Sub=2)}} \\
 \cline{2-9}
 & \textbf{\mbox{HC~data} (mean$\pm$std)} & \textbf{\mbox{SZ~data} (mean$\pm$std)} & \textbf{P\textsubscript{MW}} & \textbf{P\textsubscript{BOOT}} & \textbf{P\textsubscript{MW} (mean$\pm$std)} & \textbf{P\textsubscript{BOOT} (mean$\pm$std)} & \textbf{P\textsubscript{MW} (mean$\pm$std)} & \textbf{P\textsubscript{BOOT} (mean$\pm$std)} \\\hline
\textbf{Degree (Density)} & 71.746$\pm$2.178 & 69.207$\pm$5.149 & 0.130 & 0.086 & 0.146$\pm$0.05 & 0.089$\pm$0.044 & 0.162$\pm$0.079 & 0.103$\pm$0.063 \\
\textbf{Strength} & 37.053$\pm$7.362 & 31.680$\pm$10.246 & 0.124 & 0.079 & 0.143$\pm$0.047 & 0.096$\pm$0.029 & 0.155$\pm$0.071 & 0.108$\pm$0.045 \\
\textbf{Bin CC} & 0.986$\pm$0.025 & 0.960$\pm$0.057 & 0.150 & 0.124 & 0.150$\pm$0.052 & 0.130$\pm$0.038 & 0.184$\pm$0.089 & 0.138$\pm$0.057 \\
\textbf{Wei CC} & 0.489$\pm$0.105 & 0.415$\pm$0.144 & 0.150 & 0.105 & 0.155$\pm$0.052 & 0.119$\pm$0.027 & 0.183$\pm$0.082 & 0.128$\pm$0.050 \\
\textbf{Bin CPL} & 1.017$\pm$0.030 & 1.052$\pm$0.071 & 0.130 & 0.084 & 0.153$\pm$0.052 & 0.089$\pm$0.044 & 0.162$\pm$0.079 & 0.102$\pm$0.063 \\
\textbf{Wei CPL} & 2.216$\pm$0.610 & 2.618$\pm$0.892 & 0.124 & 0.129 & 0.151$\pm$0.061 & 0.150$\pm$0.054 & 0.155$\pm$0.072 & 0.159$\pm$0.072 \\
\textbf{Bin Eff} & 0.991$\pm$0.015 & 0.974$\pm$0.035 & 0.130 & 0.086 & 0.150$\pm$0.052 & 0.089$\pm$0.044 & 0.162$\pm$0.079 & 0.102$\pm$0.063 \\
\textbf{Wei Eff} & 0.525$\pm$0.085 & 0.466$\pm$0.113 & 0.113 & 0.078 & 0.147$\pm$0.051 & 0.099$\pm$0.023 & 0.143$\pm$0.065 & 0.111$\pm$0.041 \\
\textbf{Modularity} & 0.004$\pm$0.009 & 0.016$\pm$0.026 & 0.043 & 0.133 & 0.053$\pm$0.018 & 0.128$\pm$0.019 & 0.059$\pm$0.030 & 0.144$\pm$0.036 \\
\textbf{Coreness} & 0.018$\pm$0.022 & 0.037$\pm$0.036 & 0.164 & 0.122 & 0.141$\pm$0.060 & 0.122$\pm$0.039 & 0.199$\pm$0.091 & 0.138$\pm$0.061 \\
\textbf{Small-Worldness} & 1.003$\pm$0.005 & 1.015$\pm$0.019 & 0.178 & 0.085 & 0.146$\pm$0.062 & 0.090$\pm$0.013 & 0.214$\pm$0.098 & 0.096$\pm$0.030 \\
\end{tabular} }\vspace*{1.5mm} \\
\footnotesize{{The global version of the degree represents the mean of all edges associated to all nodes, whereas the density is the ratio between the number of present edges and the number of possible edges in the network. For this reason, the result coinci- \\  des, and it is reported only once. HC: Healthy Controls; SZ:$\,$ Schizophrenic Patients;$\,$ RST: Robust Statistical Test; MW: Mann Whitney Test; BOOT: Bootstrap Hypothesis Test; \#Sub: number of subjects removed; CC: Clustering Coefficient; \\ \hspace*{-9cm} CPL: Characteristic Path Length; Eff: Efficiency.{\label{table}}%
}}
\end{table}

\begin{figure}[!h]
\vspace*{-1mm}
\centering
\includegraphics[width=0.96\textwidth]{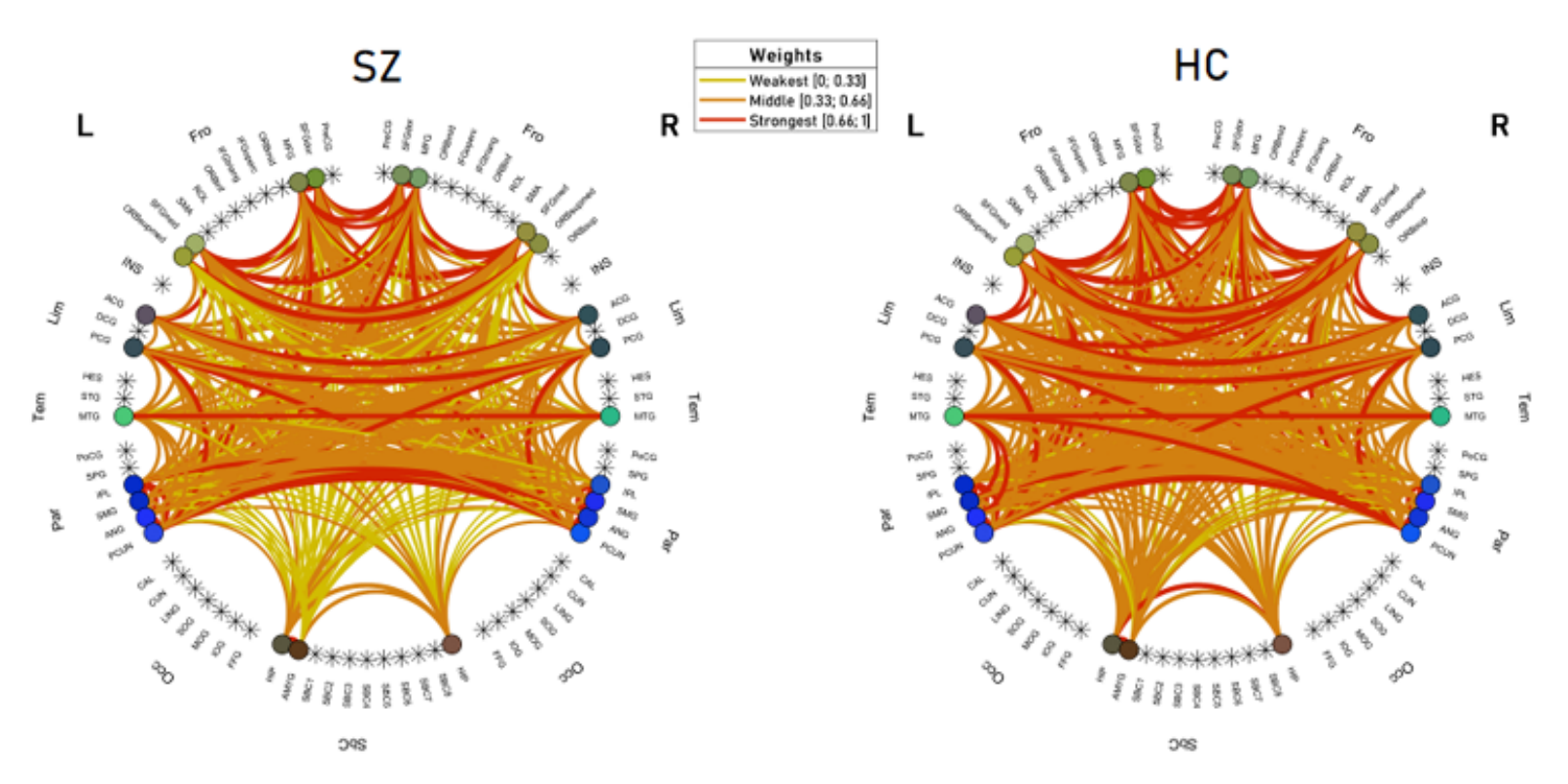}\vspace*{-4mm}
\captionsetup{justification=justified}
\caption{{Connectograms composed by only nodes of the DMN and positive connections between pairs of these nodes.{\label{fig4}}%
}}

\vspace*{4mm}
\centering
\includegraphics[width=0.96\textwidth]{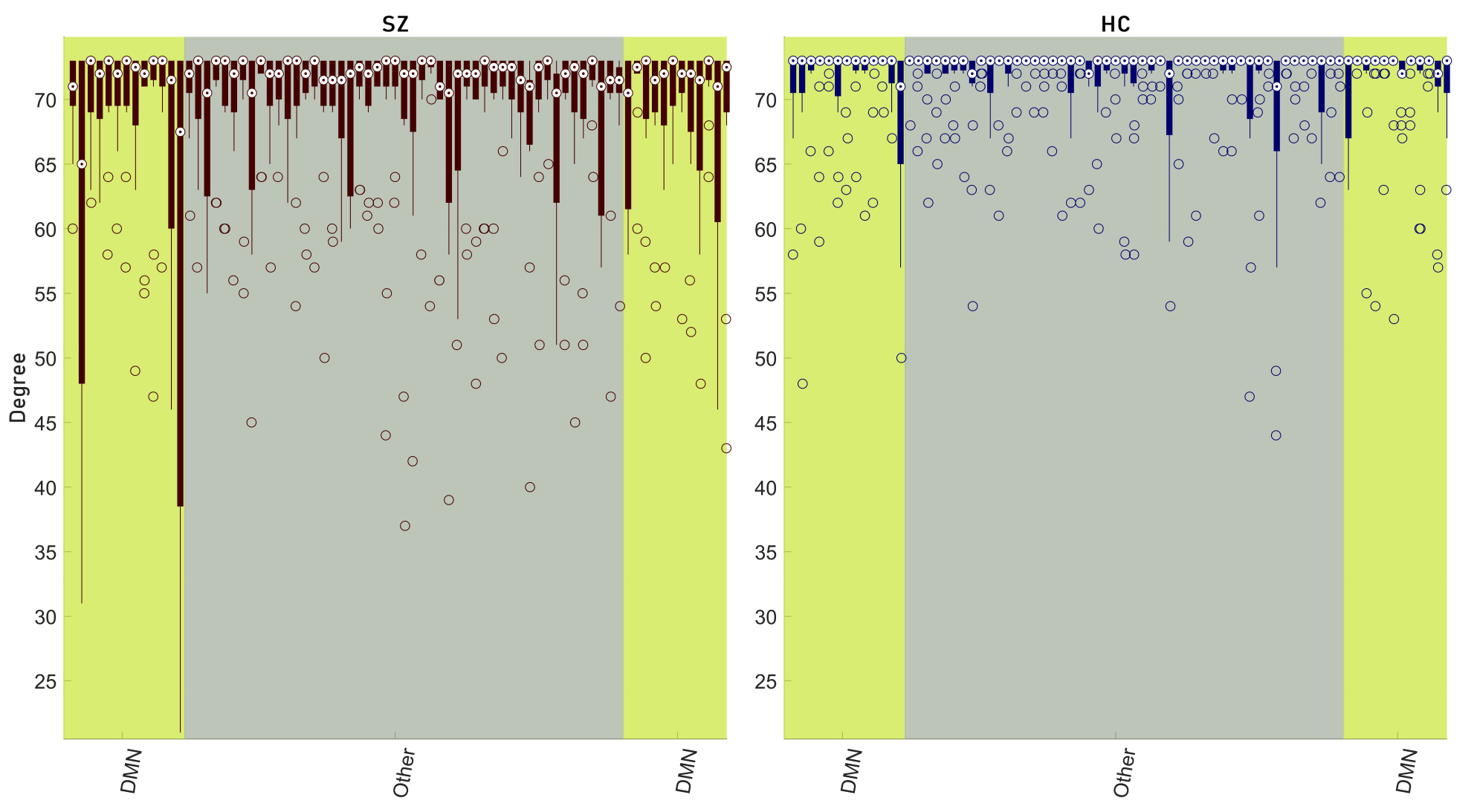}
\captionsetup{justification=justified}
\caption{\small{Distributions of the local degree values in the two populations divided according to nodes of
DMN in left and right hemisphere (yellow) or not (gray).{\label{fig5}}%
}}
\end{figure}

It is possible to notice that HC presented increased values for all segregation indexes and efficiency. Conversely, concerning the characteristic path length, mesoscale analysis, and small-worldness SZ resulted to have higher values. Differences are then enhanced in the comparison between MW and BOOT. In particular, higher p-values in BOOT are found for the weighted characteristic path length and modularity indexes. Specifically, the latter highlighted a statistically significant ($p<0.05$) difference between HC and SZ modularity that was not confirmed by using BOOT.
The robustness assessment highlighted more stable results for BOOT. Indeed, it is worth noting that the mean of the p-values, considering all indexes and removing both one and two subjects, remains closer with BOOT than the MW test in almost all indexes considered. In few cases (binary and weighted clustering coefficient in RST1) the difference between the mean of the robust assessment tests and initial tests is slightly minor in MW or comparable. Furthermore, the variability was assessed, resulting in lower standard deviation employing BOOT with respect to MW in almost all indexes. In few cases (e.g. weighted characteristic path length and modularity in RST2) this reduction is not enhanced with comparable values.
Second, DMN was extracted from all subjects of the two populations. In Figure~\ref{fig4} the connectograms of this sub-network obtained averaging all positive connections across the whole populations are shown. It is worth noting that from a visual inspection a greater number of weak connections is highlighted in the SZ group, represented by the yellow edges in Figure~\ref{fig4}.

Second, we analyzed the local values of the indexes in the DMN and across the lobes, showing the distributions of the values of the degree across the same group in Figure~\ref{fig5}. Differences between DMN regions and the other areas in SZ and HC, because of the different influence of the negative connections, were highlighted. Specifically, the most variable indexes resulted to be located within the DMN for the SZ group. On the other hand, they are more distributed across all the brain regions in HC group, even resulting more variable in the areas not included in the DMN. More precisely, one of the most variable values of the degree in both groups was the one assumed by the amygdala, located at the end of the first yellow box in Figure~\ref{fig5}, representing the left hemisphere nodes of the DMN.

\begin{figure}[!b]
\centering
\includegraphics[width=0.96\textwidth]{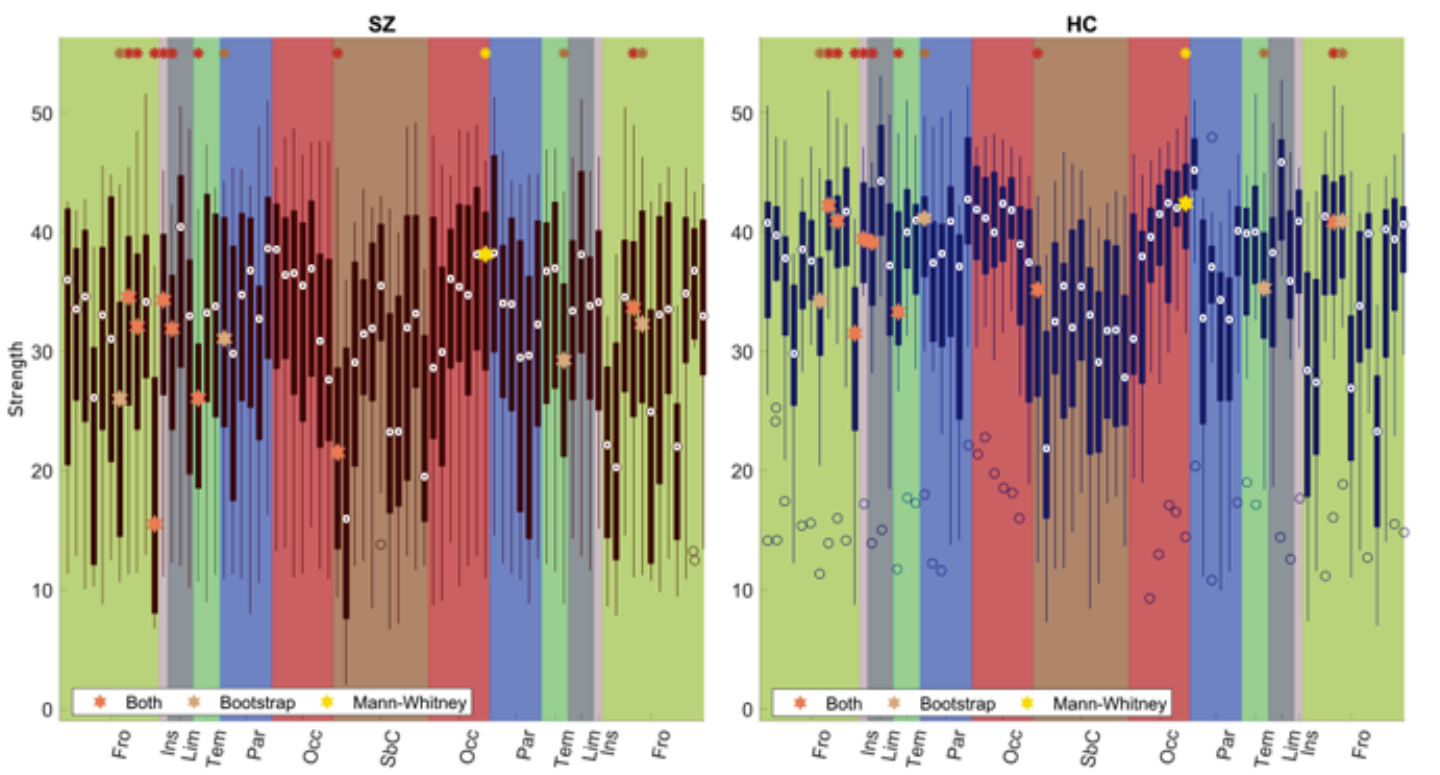}
\captionsetup{justification=justified}
\caption{{Distributions of the local strength values in the two populations divided according to the lobes.
The order and colors of the lobes is the same as the connectograms (Figures~\ref{fig4} and~\ref{fig7}) from the left to the right hemisphere. Strength of nodes which are statistically significant different ($p<0.05$) between the two populations are shown in different colors according to the different methods used: pale brown for bootstrap hypothesis testing, yellow for Mann-Whitney and red for statistically significantly different nodes found by both. More specifically, both methods identify differences in the left Rolandic operculum, supplementary motor area, medial orbital superior frontal gyrus, insula, anterior cingulate and paracingulate gyri, hippocampus, Heschl’s gyrus and right supplementary motor area, bootstrap hypothesis testing in left orbital part of the inferior frontal gyrus, postcentral gyrus, right Rolandic operculum and right Heschl’s gyrus and Mann-Whitney test in the right calcarine fissure and surrounding cortex.{\label{fig6}}%
}}
\end{figure}

Afterwards, the strength was analyzed, and the resulting distributions are shown in Figure~\ref{fig6}. First, it was noticed that the distributions of the SZ have a slightly higher variability in the majority of the nodes with respect to HC. More specifically, the most variable local indexes among their populations were extracted from frontal, parietal lobes and subcortical structures. On the other hand, general slightly higher values of strength are present in the HC group. Furthermore, the distributions of the strengths are shown in Figure~\ref{fig6} highlighting statistically significant differences ($p<0.05$) found through both MW and BOOT (in red). Most of the differences were found in regions of the frontal lobe of both hemispheres. Significance of BOOT only (in pale brown) resulted in the left orbital part of inferior frontal gyrus, postcentral gyrus, Heschl’s gyrus, and right Rolandic operculum nodes. Significance of MW only (in yellow) was found in the in right calcarine fissure and the surrounding cortex node. The complete list of the significant different node strengths is reported in the caption of Figure~\ref{fig6}.

\begin{figure}[ht]
\vspace*{-3mm}
\centering
\includegraphics[width=0.98\textwidth]{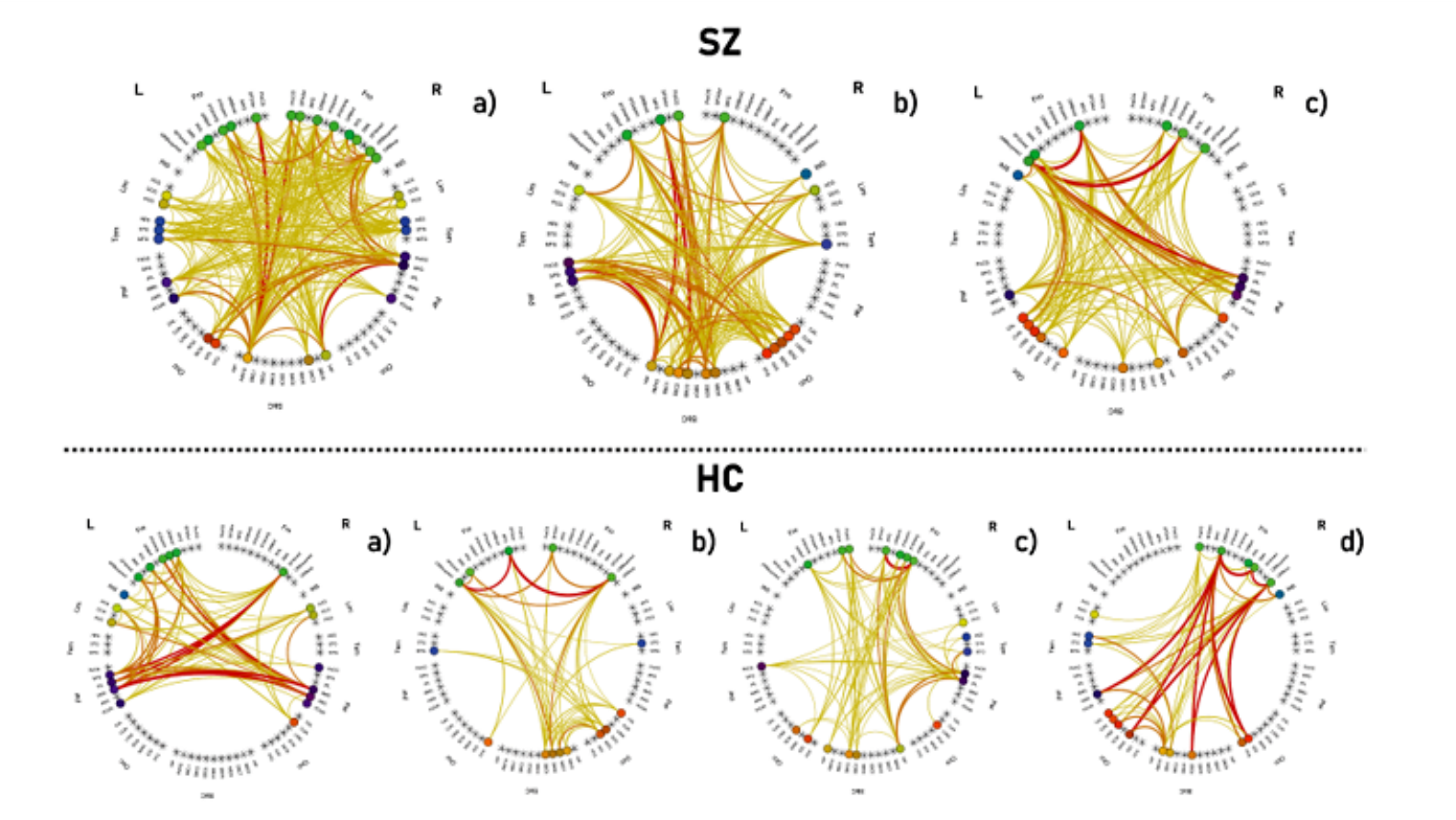}\vspace*{-2mm}
\captionsetup{justification=justified}
\caption{{Connectograms of the communities detected in the negative average group networks. On top, the three communities found in the SZ group (modularity=0.079), whereas, on bottom, the four communities of HC group (modularity=0.165). The edges are color-coded in red, orange, and yellow to represent the strongest, middle, and weakest negative connections within the community, respectively. No thresholding was applied.{\label{fig7}}%
}}
\end{figure}

Third, we analyzed the negative connections from the group average matrixes. Results from community detection analysis are shown in Figure~\ref{fig7}. Three communities, a) b) and c) on top of the Figure, composed of 32, 22, 20 brain regions respectively were identified for the negative SZ group. Four communities, a) b) c) and d) on bottom, composed of 20, 15, 20 and 19 regions respectively were identified for the negative HC group. A difference between the formed communities is the inclusion of most frontal, limbic and temporal regions in the community a) of SZ group, which is not observable in any HC community. Furthermore, the configuration of parietal regions from both hemispheres, mostly included in the community a) in HC, is not noticeable in SZ, where these regions are divided according to hemisphere b) and c) of SZ, also appearing to be clustered with the contralateral occipital lobe.

\begin{figure}[!h]
\vspace*{-1mm}
\centering
\includegraphics[width=0.59\textwidth]{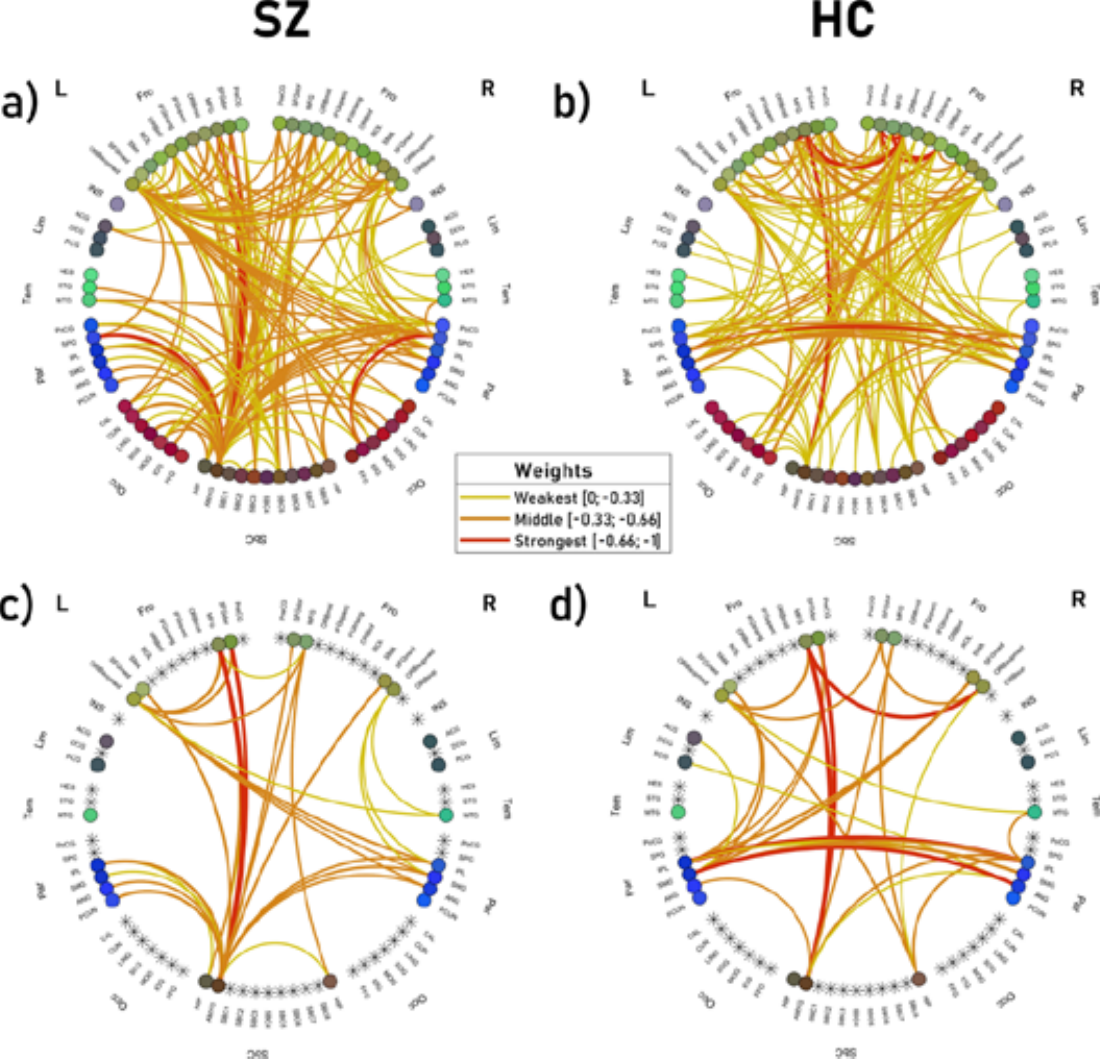}
\captionsetup{justification=justified}
\caption{\small{Connectograms of negative average group networks. a) and b) are the results on the whole-brain, c) and d) on only the DMN of SZ and HC groups respectively. 5\% density thresholding is applied on the shown connectograms for graphical clarity only.{\label{fig8}}%
}}\vspace*{-3mm}
\end{figure}

Then we visualize the strongest and frequent negative edges in the whole brain and in the DMN, as shown in Figure~\ref{fig8}. Specifically, the connectograms formed by the average of all matrixes composed by the negative edges in the whole brain (7a and 7b) and in the DMN (7c and 7d) are shown applying a threshold to keep 5\% strongest negative edges. First, SZ average network resulted to have more negative connections in the whole brain with respect to HC. Another great difference noticed both in the number of preserved negative edges and in their weights was in the inter-hemispheric connectivity of the frontal lobe, as well as the parietal. In addition, it is possible to notice in SZ a greater number of negative connections in the occipital lobe with respect to HC. The configurations given by the only strongest connections in DMN highlighted similarities and differences. Among all, connections between amygdala and some frontal regions were in common, whereas strong edges between the parietal lobes of the two hemispheres were found only in HC. In SZ parietal regions are instead connected to amygdala and superior frontal gyrus. Finally, HC average matrix revealed an important edge between the latter, but in the right hemisphere, and the middle frontal gyrus, that was not found in SZ.

\section{Discussion}\label{discussion}

In this work, we proposed a multi-level approach for the case-control analysis of connectivity data structured on different levels of investigation. In particular, the assessment of the connectivity from global indexes to single edges (brain activations/deactivations) was carried out on a population \linebreak characterized by chronic schizophrenia, thought to cause both diffused and focalized dysconnectivity patterns. The multi-level analysis resulted to potentially favor more robust results in contexts where the focus is not well-known and statistical tests can be easily biased by uncertainty.

First, global-level results were analyzed. As previously mentioned, this level was examined according to three main investigations. All experiments comprised the evaluation of 11 graph-based indexes, either binary or weighted, which are amongst the most widespread in connectivity studies \cite{Rubinov2011Weight-conserving, Blasi2020Early, Bassett2017Small-World, Oliver2019Quantifying}. The first analysis involved the comparison between HC and SZ groups, compared through two statistical tests (MW vs BOOT). In Table 1 we showed the values of the indexes along with the p-values obtained from both tests. The increased segregation indexes and efficiency in HC with respect to SZ is in line to what reported from previous studies \cite{Lynall2010Functional, Liu2008Disrupted}, as well as reduced path length, mesoscale analysis and small-worldness. More specifically, MW highlighted one statistically significant difference between the two population in the modularity ($p\textsubscript{MW} = 0.043$). The modularity resulted to be significantly higher in SZ providing evidence that the deletion of negative connections, in this case analyzed separately, tend to create families in SZ that in HC are less visible. However, it is worth noting that similar studies presented different results for this particular index and also uncertainty for clustering coefficient and characteristic path length, indicating either increased or reduced values in SZ \cite{Fornito2012Schizophrenia, Liu2008Disrupted, Alexander-Bloch2010Disrupted, Avery2018Hippocampal, Berg2012Fragmentation:, Heuvel2014Brain, Yu2011Altered, Wang2010Impaired}, probably dependent on the thresholding method. For these reasons, as previously mentioned, the reliability of these results was assessed through BOOT because of the reduced size of the dataset, the presence of the uncertainty due to the non-stationarity of the functional connectivity and the possible presence of spurious connections. The results of the bootstrap hypothesis testing highlights differences with respect to the standard nonparametric approach. On one hand, the modularity did not present significant difference ($p\textsubscript{BOOT} = 0.133$). On the other hand, a trend towards lower p-values can be noticed for all other indexes, except the weighted path length ($p\textsubscript{MW}=0.124$; $p\textsubscript{BOOT}=0.129$) and the above-mentioned modularity. One of the most replicated results is the significant difference found in the efficiency between HC and SZ. According to different studies, functional connectivity fluctuations appear to be coordinated throughout the brain so as to realize global variations in network efficiency, which could represent a balance between optimizing information processing and minimizing metabolic expenditure \cite{Fornito2012Schizophrenia, Sun2019Dynamic, Zalesky2014Time-resolved}. Furthermore, examining studies with a greater number of subjects \cite{Unschuld2014Prefrontal}, or of dynamic connectivity \cite{Sun2019Dynamic}, the results obtained through BOOT seem to be closer than MW. At this stage of understanding, we may speculate that BOOT seems to provide an alternative method to test more robustly the differences between functional connectivity of two groups, especially when composed by low number of data and subjects.

However, we acknowledge that there is considerable debate among researchers as to uncertainty influence and integration of dynamical functional connectivity data. For this reason, we extended the analysis comparing MW and BOOT through robustness assessment tests. In particular, the variability of the statistical estimate was evaluated at the removal of one or two subjects from the whole population, just as in leave-n-subject-out cross validation. In Table 1 the results of these experiments are summarized, finding clear support for the hypothesis of more stability and reliability of the BOOT results. In particular, the index mean values obtained by BOOT across the random extractions (27 for leave one out and 350 for leave two out) of subjects to be removed remain much more stable with respect to the values obtained with entire sample (e.g., degree - RST1: $p{\textsubscript{MW}=0.130}$; $mean(p\textsubscript{MW})=0.146$; $p\textsubscript{BOOT}=0.086$; $mean(p\textsubscript{BOOT})=0.089$; RST2: $mean(p\textsubscript{MW})=0.162$; $mean(p\textsubscript{BOOT})=0.103$).  Also, the standard deviation across all iterations resulted to be much lower by using BOOT than MW (e.g., small-worldness - RST1: $std(p\textsubscript{MW})=0.062$; $std(p\textsubscript{BOOT})=0.013$; RST2: $std(p\textsubscript{MW})=0.098$; $std(p\textsubscript{BOOT})=0.030)$. The observations regarding mean and standard deviation can be generalized to all indexes considered, resulting in less variable results or, in few cases remaining similar to MW. More specifically, the cases of the resulting mean of binary and weighted clustering coefficient for RST1, or the standard deviation of the weighted characteristic path length and modularity for RST2, which anyway did not result to be less stable with BOOT.  In general, the results highlighted a higher stability and robustness of the BOOT in RST1 and RST2,  providing support to its possible use when dealing with a small dataset which is also affected by uncertainty of the measures.

It is worth discussing that the results revealed by the robustness tests may answer the question of evaluating the results of the functional connectivity in a more reliable way. Although the needs for dynamic functional acquisition \cite{Hutchison2013Dynamic, Hindriks2016Can, Damaraju2014Dynamic} and of more attractive methods for removing spurious connections are well-known \cite{Fornito2012Schizophrenia, Civier2019Is, Hagmann2008Mapping, Tewarie2015minimum}, having a reliable method to evaluate functional connectivity of small datasets, which can potentially approximate to results from greater datasets and avoid the loss of information, can help in the interpretation and understanding.

Second, focusing on the results of the second level, comprising sub-network (DMN) and local indexes, interesting findings were noticed. It is worth noting that a first visual difference between the mean matrix of the groups was visible. Indeed, in Figure~\ref{fig4}, the connectograms highlight weaker connections in the DMN of SZ with respect to HC, in agreement to previous studies \cite{Hu2017Review, Unschuld2014Prefrontal}. However, no significant differences between the indexes of the sub-networks were found. On the other hand, the evaluation of the local indexes allowed to identify some notable patterns.
From Figure~\ref{fig5}, it is indeed possible to notice that the most varying local degrees are those referred to nodes of the DMN, especially in the SZ group. Analyzing, instead, the results in Figure~\ref{fig6} on the strength index, the differences are less visible although a slightly greater value is noticed across all nodes of the HC group, together with a confirmed greater variability distributed across the whole brain in the SZ group. On one hand, these findings report a greater presence of negative connections in the DMN of SZ, which can be seen as an inverted connectivity nature between prefronto-temporal areas, in agreement to diversity found in a previous study \cite{Bassett2012Altered}. This is particularly indicative considering the involvement of the DMN in SZ \cite{Hu2017Review} and that many symptoms of the pathology mirrors a failure in the integration between the generated behaviour and concurrent perceptive phenomena \cite{Hemsley2019Perceptual}. On the other hand, this variability can also be influenced by the limitations of the data. It is indeed well-known as one of the states identified by performing dynamical connectivity experiments can be traced to the  DMN \cite{Damaraju2014Dynamic}.

Furthermore, we analyzed the nodes which had the greatest variability of the degree index and that mostly differed in the strength value. Amygdala, medial orbital part of the superior and inferior frontal gyrus and part of the cingulate cortex resulted to be the nodes which were most affected by the presence of negative connections in the SZ group. Interestingly, amygdala, which is also a well-known hub of the brain \cite{Tomasi2011Association}, part of the superior frontal gyrus and cingulate cortex are all part of the DMN, whereas the inferior frontal gyrus not. However, the inferior frontal gyrus was widely studied because of its divergent characteristics of anatomical/functional connectivity in SZ and its relation to semantic processing \cite{Jeong2009Functional}. The enhanced variability may be indeed related to specific characteristics and deficiencies of the patients, which may be potentially investigated through correlation studies with demographic and severity variables, or, in this case, semantic function tests, as already widely done in literature \cite{Unschuld2014Prefrontal, Blasi2020Early, Yu2011Altered}. Afterwards, we also computed MW and BOOT tests on the local values of the strength, highlighting the nodes mostly differing in the value of this index between the two populations. In general, it is possible to notice that the most different local indexes are those referred to frontal, parietal lobes and subcortical structures, in agreement to previous study \cite{Zalesky2010Network-based, Alves2019improved, Sormaz2018Default}. In particular, significance of left orbital part of inferior frontal gyrus, postcentral gyrus, Heschl’s gyrus and right Rolandic operculum nodes were found by BOOT and right calcarine fissure and surrounding cortex node by MW. The results on these areas appear to be in line with previous studies, where these nodes were found to be different, although the case of Heschl’s and postcentral gyri were found in the specific case of SZ patients with auditory hallucinations \cite{Yu2011Altered, Shinn2013Functional, Koshiyama2020Neurophysiologic}.

The community detection analysis of the anticorrelations highlighted differences between the two groups. First, in SZ, most nodes from frontal, limbic and temporal lobes are clustered together in the first community. This deactivation organization may be relevant in the study of failed activity inhibition of schizophrenia characterized by selective disruption of an automatic inhibitory process, and failure to limit the current contents of consciousness \cite{Peskin2020Subcortical, Beech1989Evidence}. In the parietal lobes configuration other differences were noticed, especially regarding the clustering with contralateral occipital lobe in SZ. In other studies \cite{Henseler2010Disturbed, Fujimoto2013Dysfunctional, Yildiz2011Parietal}, these lobes were investigated in relation to SZ highlighting important functions. For example, it was hypothesized that cognitive deficits and delusions may be related to malfunctions in the parietal lobe \cite{Yildiz2011Parietal} or that the maintenance of visuospatial information is associated to a network of occipital cortex regions \cite{Henseler2010Disturbed}. It would be of great interest to study correlations between symptoms and activity in these communities, as the contralateral negative cooperation may reinforce the hypothesis of an involvement of parieto-occipital sub-network in auditory hallucination, such as the positive correlation found in an EEG and MEG study \cite{Fujimoto2013Dysfunctional}, or the relations to other disorganization symptoms.

Third, the connectogram pattern of negative edges obtained by averaging all the negative contributions in the whole brain and DMN of the groups highlighted interesting indications. Primarily, SZ resulted to have more and strongest negative connections in the whole brain with respect to HC. Second, an increased deactivation in the occipital lobe appear to confirm the results previously obtained through network based statistics, which found dysconnections in the same areas \cite{Zalesky2010Network-based}. Then, the number and the weight of the preserved negative correlations are generally greater in SZ in the frontal, which includes prefrontal as well, and parietal lobes, in agreement to what previously shown on sub-network and nodes. This finding appears to be in line with previous studies where different organization of the connectivity was found in prefrontal and frontal lobes \cite{Pettersson-Yeo2011Dysconnectivity, Zhou2018Altered}. As said, the anticorrelations could indicate an abnormal inhibition of some regions or lobes activity. In the context of DMN, except the connections from amygdala to the frontal regions expected to be common to both groups being a well-known hub of the brain networks, as abovementioned \cite{Tomasi2011Association}, a different configuration of the edges was noticed. First, in the HC group, the parietal lobes of the two hemispheres are deactivated with strong interactions that are not present in the SZ group, in agreement to community detection analysis. In this case, parietal regions of DMN mostly communicate with amygdala and superior frontal gyrus. Furthermore, HC group is characterized by a very strong negative connection between middle frontal gyrus and the medial orbital part of the superior frontal gyrus that is not even present in the SZ group. These findings may be related to significant small-world topology abnormalities found in regions of the prefrontal and parietal lobes in a previous study \cite{Liu2008Disrupted}. Indeed, the difference of negative network involving the DMN regions of the parietal lobe is the most evident, both for the number of edges and their interactions with the DMN  regions of the frontal lobe. From previous works \cite{Hu2017Review, Fox2005human, Fransson2005Spontaneous}, it is known as intrinsic task-positive and task-negative networks exist in HC during rest, comprising some areas of these lobes. It can be speculated that this organization may be responsible for external environmental and self-referential processing respectively and serves mental processes. In individuals with SZ or paranoid tendencies, abnormal negative connectivity within and between these two networks could cause them to be oversensitive to both internal thoughts and external stimuli. In this sense, abnormal presence of anticorrelations, together with a low integration between the two networks, may reflect abnormal inhibition of the other circuit and that extrospective and introspective thinking is switched back and forth in an overly excessive manner \cite{Hu2017Review, Zhou2007Functional}. In this work, differences were also found in the temporal lobe that, however, in deactivation analysis of our dataset remained comparable. Minor differences given by the presence of some negative weak connections in HC which are not present in SZ were found in the regions of the DMN of the limbic lobe, confirming the affection of the functional reorganization of the SZ originating from posterior cingulate cortex \cite{Skudlarski2010Brain}. This resulted to be in line with previous findings on cingulate gyrus in SZ and its importance in functional integration \cite{Segal2010Diffusion}.

However, it is worth noting that the small size of the HC and SZ groups available for this study is a limitation. A subsequent validation of the bootstrap method on graph-based indexes obtained from a wider functional connectivity dataset would be attractive. Although the greater stability to changes of the measures and a trend towards results of larger studies is promising, validation with more subjects is needed.  A related limitation concerns the lack of discrimination among different types and severity levels of schizophrenia \cite{Hu2017Review}. Also, the leave-k-subject-out procedure to test the statistical analyses may be performed on dynamical connectivity data. Thus, the assessment of the influence of the different states which are assumed during a resting-state fMRI would reinforce the potential of the approach in underpinning usability and give better understanding of the functional activations and deactivations.

Second, one of the main aims of the study was the validation of the approach through bootstrapping of the indexes accounting all connections, which include uncertainty as well, but without any loss of information or arbitrary thresholding of spurious edges and very low correlations. The results obtained through BOOT were more stable in comparison to standard nonparametric tests, being robust to noise, but at the same time it is paramount to focus and apply new and different thresholding methods to assess their influence, as briefly mentioned. Examples are percolation or spanning tree approaches \cite{Fornito2012Schizophrenia, Alexander-Bloch2010Disrupted, Civier2019Is, Hagmann2008Mapping, Tewarie2015minimum} appearing to be less arbitrary than typical density or absolute thresholding and which may potentially improve even more the estimates. The former, for example, was widely employed in a number of studies to find a potential optimal deletion of the spurious edges maintaining brain network structure and connectedness \cite{Bordier2017Graph, Nicolini2017Community, Bordier2018Disrupted, Guo2021Percolation}. In this perspective, results from bootstrap analysis may be investigated in relation to percolation analysis. Another possibility could be to apply bootstrapping of the groups at connection-level to obtain more robust templates for comparison, identify errors and quantify the uncertainty of the measure.

Third, the top-down approach was implemented in the SPIDER-NET tool which automatically allows the investigation of topological properties that are the most widespread in brain connectivity studies \cite{Coluzzi2022Development}. However, some of the indexes analyzed have intrinsic limitations, such as the weakness of Newman’s modularity to detect small modules [84]. In addition, thresholding methods can strongly affect the results of community analyses and topological indexes. For this reason, further investigations and comparisons to other methods, such as random walk-based approaches for community detection, would be of great interest.

Finally, the robust multi-level approach helped to identify differences in the connectivity of SZ patients. More specifically, it did not result in increased or decreased connectivity with respect to HC but in a different configuration, highlighted by the bootstrapping. As said, the most replicated graph-feature is the efficiency, whereas other features appear to be more affected by the pre-processing and there is no strict agreement. Also, the great intra-group variability, particularly relevant for SZ, plays an important role that may be investigated in future considering the clinical condition and severity of the patients or other specific characteristics and deficits. From these considerations, the results confirmed that the connectivity in SZ is not systematically increased/decreased but generally different, as already reported in another study \cite{Skudlarski2010Brain}. Nonetheless, it is clear and replicated that some areas are affected more in SZ \cite{Fornito2012Schizophrenia, Heuvel2014Brain}. However, it remains unclear whether SZ abnormalities are the result of a localized dysconnection exerting widespread effects throughout the brain, or a whole-brain dysfunction that affects certain regions more. New investigations would be certainly needed to further investigate the causes and effects related to the onset of the condition.

\section{Conclusion}\label{conclusion}
In this work, we described a multi-level robust approach for the investigation of the brain connectivity. Starting from the measures of global connectivity, which can be robustly and more stably explored through bootstrapping, general abnormalities in case-control groups can be identified. In some cases where the connectivity is not evidently increased/reduced, or particularly different in some areas, going towards lower levels can be helpful. In this case, we analyzed separately the functional connectivity in the DMN, beyond those indexes related to specific nodes. Finally, the activations and the deactivations were investigated. The whole procedure is developed in the SPIDER-NET software tool (https://caditer.dongnocchi.it/spidernet/) to automatically help in the different level investigations of various pathology to better understand their nature and effects.

\end{document}